\documentclass[prd, showpacs,nofootinbib,amsmath,amssymb]{revtex4}
\usepackage{amsfonts, amssymb, amsmath, graphicx, comment, bm, slashed}
\usepackage[colorlinks]{hyperref}
\usepackage{dcolumn}   
\usepackage{bm}        
\usepackage{subfig}
\usepackage{caption}
\usepackage{multirow}
\usepackage{mathrsfs}

\begin{document}
\title{Magnetic-induced condensate, vortices and vortons in\\
	 color-flavor-locked-type matter}
\author{Fu-Ping Peng$^1$}
\author{Xiao-Bing Zhang$^1$}
\author{Yi Zhang$^2$}
\affiliation{$^1$School of Physics, Nankai University,\\Tianjin  300071, China}
\affiliation{$^2$ Department of Physics, Shanghai Normal University,\\ Shanghai 200230, China}

\date{\today}

\begin{abstract}
By considering Higgs modes within the Ginzburg-Landau framework, we study influences of a rotated magnetic field on
the color-flavor-locked-type matter of dense QCD. We demonstrate, in a model-independent way, that a diquark condensate
may be triggered by the magnetic response of rotated-charged Higgs modes, in addition to the known color-flavor-locked 
condensate.
Moreover, the condensate is applied to explore formations of vortices in the presence of external magnetic fields. The superfluid-like vortices are constructed for the magnetic-induced condensate. In the situation including both kinds of condensates, the theoretical possibility of vortons is suggested and the formation condition and the energy stability are investigated semi-classically.

\end{abstract}

\maketitle

\section{Introduction}

Strongly interacting matter under the influence of magnetic fields has attracted intensive interests
in recent years, due to the realistic relevance to phenomenology in relativistic heavy-ion collisions
and astrophysical objects in the universe, for recent reviews, see, e.g., ~\cite{andersen2016phase,kharzeev2013strongly}.
In heavy ion collision experiments, a strong magnetic field up to $eB \sim 2m_{\pi}^2$ with $m_{\pi}$
the pion mass, i.e., $B \sim 10^{18}\text{G}$ can be produced in non-central collisions at Relativistic Heavy
Ion Collider (RHIC), and $B \sim 10^{20}\text{G}$ at Large Hadron Collider (LHC)~\cite{kharzeev2008,skokov2009}.
For astrophysics objects, the magnetic fields as large as
$B \sim 10^{14}$ - $10^{15}\text{G}$ exist on the surface of Magnetars. While the strength
can reach $B \sim 10^{18}\text{G}$ in the interior of regular neutron stars, a theoretical upper limit to the magnetic field may stand as high as $B \sim 10^{20}\text{G}$ inside self-bound compact stars~\cite{dong2001,lai1991cold}.
Under such circumstances of strong magnetic fields,
answer to the question of how the Quantum chromodynamics (QCD) phase structure can be modified remains
one of the major theoretical challenges. For instance, at vanishing chemical potential, a general effect
of magnetic field on the vacuum structure of QCD is an enhancement of the dynamical symmetry breaking,
a phenomenon usually referred to as ``magnetic catalysis''~\cite{miransky2015quantum}. At zero
temperature, recent lattice QCD study of the behavior of the $u$- and $d$-quark condensates in
magnetic fields has confirmed the magnetic catalysis phenomena, however, for temperatures of the order of
the crossover temperature, a decrease of the quark condensates is found, this is so called ``inverse
magnetic catalysis'' effect at certain values of the magnetic
field~\cite{bali2012,miransky2002,fukushima2012}.
Up to date, the mapping of QCD phase diagram onto the temperature and 
magnetic field plane has not been firmly established.

Further, through intensive studies in past decades, it is realized that there exists so-called color
superconducting phase of quark matter at high density regime (see, e.g.
Refs.~\cite{alford2004dense,buballa2005njl}). In the scope of color superconducting matter, it is of
great interest to investigate which modifications are induced by the magnetic-field background.
Besides its theoretical implication to the QCD phase diagram, this topic might be very important for the physics of magnetars since 
the inner region of compact stars is suggested to consist of color superconductors.
As the most typical color superconductor, the color-flavor-locked (CFL) phase is widely believed to be
the ground state of QCD at extremely high density and low temperature~\cite{alford1998qcd}. Due to the
diquark condensates formed in the CFL phase, the symmetry breaking pattern is
\begin{eqnarray}
G=SU(3)_{C}\times SU(3)_{L}
\times SU(3)_{R}\times U(1)_{B} \rightarrow H=SU(3)_{C+L+R}\equiv SU(3)_{C+F},\label{cfl}
\end{eqnarray}
where the approximate symmetry $U(1)_{A}$ in $G$ and the discrete symmetries in $H$ have been ignored.
It means that an original QCD symmetry, including color, flavor (left- and right-hand) and baryon number
symmetries, is broken down to the color-flavor locked symmetry.
The main feature relevant for magnetic effect is the rotated electromagnetic mechanism, namely an unbroken
$U(1)_{\widetilde{Q}}$ group is embedded in the color-flavor locked subgroup where the rotated electric
charge is defined by $\widetilde{Q}=Q_{F}\times {1}_{C}-{1}_{F}\times Q_{C}$ in the color-flavor space.
As the spin-$0$ color superconductor, the diquark condensates in CFL are always neutral in the sense of
rotated charge.
Since CFL is not electromagnetic superconductor, there is no Meissner effect and the unscreened magnetic
fields propagate unscreened inside the color-superconducting matter~\cite{alford1998qcd,alford2000magnetic}.
In the situation of strong gauge coupling, the rotated electromagnetic field, as the combination of
electromagnetic and gluon fields~\cite{alford1998qcd,alford2000magnetic}, is made up mostly of a usual
electromagnetic field. Thus, the rotated magnetic field can be described by an external background field
$B$ and the unit of rotated charge can be given by an electron charge $e$ approximately.

By introducing an applied field $B$, there exist considerable changes in the properties of color
superconductor. For three-flavor color-flavor-locked-type matter, in particular, the original symmetry
breaking pattern is possible to become violated. Once strong magnetic fields are introduced, the
color-flavor-locked group is broken as
\begin{eqnarray}
H\rightarrow M=SU(2)_{C+F},\label{cfl2}\end{eqnarray}
which corresponds to the magnetic color-flavor locked phase
(MCFL) \cite{ferrer2005magnetic,fukushima2008color,ferrer2006color,ferrer2007magnetic}.
In the literatures, the MCFL phase was mainly investigated by a phenomenological Nambu-Jona Lasinio (NJL)
model with four-quark interactions and many studies were devoted to the magnetic-field dependence of
color-superconducting gaps.
From the NJL numerical calculations, the gaps are observed to display de Haas-van Alphen
oscillation \cite{ferrer2005magnetic,fukushima2008color}.
The NJL analytical derivation is valid in the limit of strong magnetic field. There, it is found that
the large magnetic fields could enhance the specific gap \cite{ferrer2006color,sen2015anisotropic} which
might be considered as a generic magnetic catalysis.

Alternatively, the color-superconducting phase and the inhomogeneous topological vortices consisting of it can be studied in a model-independent Ginzburg-Landau (GL) framework, see, e.g., ~\cite{giannakis2002ginzburg,iida2002superfluid,balachandran2006semisuperfluid,nakano2008non,eto2014vortices}.
To account for the CFL phase, the GL Lagrangian can be developed from the symmetry breaking pattern Eq.~(\ref{cfl}).
The diquark-condensate order parameter is usually described by a $3\times3$ matrix, which is rather simple with
respect to a $72 \times 72$ matrix in the phenomenological model of quarks. Consequently, the relevant degree
of freedom (dof) are the Higgs modes, i.e., fluctuations of diquark condensates, rather than the microscopic
quarks and gluons dof. In a previous work~\cite{zhang2015magnetic}, the response of the Higgs modes to the
external magnetic fields was investigated. As a result, the threshold value of magnetic fields was obtained
for the transition to MCFL and the influence on the CFL vortices was considered preliminarily.

In the present work, magnetic effect will be further studied via the systematic GL analyses. Our attention focus on the case that the magnitude of magnetic field is larger than the threshold. On the one hand, the studies in this paper is going to shed light on some unknown aspects regarding magnetic response of Higgs modes.
Specifically, the additional diquark condensate caused by the response of rotated-charged Higgs modes is stressed to be necessary for color-flavor-locked breaking. In the resulting less-symmetric ground state, we attempt to explain the magnetic dependence of gap qualitatively. As well known, on the other hand, the GL theory is an useful tool for exploring the topological objects such
as the spatial-dependent vortices. In particular, the superfluid vortices and the so-called non-Abelian vortices have been suggested for
color-flavor-locked-type matter.
One must note that, the formation mechanism of these vortices are irrelevant to an external magnetic
field. Henceforth, it is one of the important issues to explore the vortices consisting of MCFL. By taking
the magnetic-induced condensate into account, we investigate two kinds of vortex solutions for different magnetic
fields and boundary conditions. Besides the superfluid-like vortices from $U(1)$ breaking, the scenario of
topologically and energetically stable vortons is suggested in the present work. These theoretical possibilities
are pointed out for the first time and they might have potential applications in the astrophysical environment.

This paper is structured as follows. After a brief review of the GL formalism accounting for CFL, in
Sec.~\ref{sec:2}, particular emphases are placed on the magnetic response of charged Higgs modes and its
consequences on the modification of the ground state. In Sec.~\ref{sec:3}, we mainly study the formation
mechanism when both kinds of vortices are included in the presence of magnetic fields. Sec.~\ref{sec:4} is
devoted to summary and discussions of some open problems.

\section{Magnetic-induced condensate}
\label{sec:2}
\subsection{GL Lagrangian with $B$-dependent coefficient}
\label{sssec:1}
%

In the absence of magnetic fields, our starting point is the most-symmetric CFL phase with a uniform
color-superconducting gap. The GL Lagrangian is developed from the symmetry breaking pattern in Eq.~(\ref{cfl}) and it is invariant
under the original symmetries $G=SU(3)_C \times SU(3)_F\times U(1)_B$. Within the GL framework, the
diquark-condensate order parameter is denoted as a complex $3\times3$ matrix $\Phi$. Under the notation
$i = 1, 2, 3 = u, d, s$ and $\alpha = 1, 2, 3 = r, g, b$, the matrix element $\Phi_{i \alpha}$ accounts
for the pairing of quarks with non-$\alpha$ colors and non-$i$ flavors \cite{iida2002superfluid}.

To quartic order in $\Phi$, the GL Lagrangian can be written as~\cite{iida2002superfluid,giannakis2002ginzburg}
\begin{equation}
\mathcal{L}=
\texttt{Tr}\left[(\nabla\Phi)^\dagger\nabla\Phi
  -\alpha\Phi^\dagger\Phi -\beta_2(\Phi^\dagger\Phi)^2\right]
-\beta_1(\texttt{Tr}[\Phi^\dagger\Phi])^2 +\cdots ,\label{gl}
\end{equation}
where the constant term for vanishing vacuum energy and the term for gauge fields have been
ignored. For the CFL phase, the vacuum expectation value (VEV) of the diquark condensate is
\begin{equation}
  \text{VEV}(\Phi)=\texttt{diag}(v,v,v) ,\label{cflground}
\end{equation}
where the diagonal elements have been assumed to be equal \cite{iida2002superfluid}. The value of
$v$ corresponds to a degenerated CFL gap and it is obtained from
\begin{equation}
  \label{eq:dvaccum}
v^2 = -\frac{\alpha}{3\beta_1+\beta_2}.
\end{equation}
Note that the coefficient $\alpha$ is responsible for the existence of color-flavor-locked condensate
and hence is always negative, while the coefficients $\beta_1$ and $\beta_2$ are positive.

Due to the Eq.~(\ref{cfl}), the Higgs modes appear as fluctuations of diquark condensate around the
CFL vacuum. Since the order parameter space is $G/H \simeq U(3)$, the Higgs modes are made up of the
singlet field $\phi$ and the octet fields $\zeta^a$ ($a = 1, 2, \cdots, 8$). Explicitly, these
collective modes can be given by perturbing the order-parameter matrix
\begin{eqnarray}
\Phi=v\textbf{1}_3+\frac{\phi+i\varphi}{\sqrt{2}}\textbf{1}_3+\frac{\zeta^a+i\chi^a}{\sqrt{2}}T^a,
\label{pert}
\end{eqnarray}
where $T^a$ is the generators of $U(3)$ with $\texttt{Tr}[T^a T^b]=\delta^{ab}$.
In Eq.~(\ref{pert}), the singlet field $\varphi$ and the octet fields $\chi^a$ correspond one to
one to the Higgs fields $\phi$ and $\zeta^a$, respectively. They belong to the pseudo Nambu-Goldstone
(NG) modes.

For our purpose, only the Higgs modes are taken as the basic dof. From the Eq.~(\ref{gl}), the Higgs
masses can be obtained,
\begin{eqnarray}
m_\phi^2=-2\alpha,\\ m_\zeta^2=4\beta_2
	v^2,\label{mhiggs}
\end{eqnarray}
which could be treated as the GL coefficients equivalently. In terms of these Higgs masses, the GL
potential may be expressed in a more proper form, i.e.
\begin{eqnarray}
\mathcal{V}_\phi=
\frac{m_\phi^2}{12v^2}(\texttt{Tr}[\Phi^\dagger\Phi-v^2])^2,
\label{glmphi}
\end{eqnarray}
and
\begin{eqnarray}
\mathcal{V}_\zeta=
\frac{m_\zeta^2}{4v^2}\texttt{Tr}\left[\left<\Phi^\dagger\Phi\right>^2\right],
\label{glzeta}
\end{eqnarray}
where the definition $\left<M\right>\equiv M-(1/N)\texttt{Tr}M$ is used for a $N\times N$ matrix $M$.
While the potential with the singlet mass accounts for the trace contribution, the traceless
contribution is encoded in the potential with the octet mass. The Higgs octet stems from the non-Abelian
feature of Eq.~(\ref{cfl}) and part of them are possible to carry the rotated charges defined by an
unbroken $U(1)_{\widetilde{Q}}$. In the presence of magnetic fields,
the properties of Higgs octet and thus the traceless potential Eq.~(\ref{glzeta}) are expected to
be influenced.

In Ref.~\cite{zhang2015magnetic}, we had considered the magnetic response of Higgs octet  and attributed it to such a change as $m_\zeta^2 \rightarrow (m_\zeta^{eff})^2$.
As long as the squared mass in Eq.~(\ref{glzeta}) is replaced by $(m_\zeta^{eff})^2$, the coefficient of traceless potential becomes magnetic-field dependent. Suppose that, except for $(m_\zeta^{eff})^2$, the formalism describing CFL remains unchanged, a GL Lagrangian with the medium-dependent coefficient can be constructed. 
It is essentially an effective theory since the CFL phase, including its vacuum, the Higgs spectrum and so on, is pre-defined while only the corrections to the known CFL results is concerned. 
Also, it was pointed out that the value of $(m_\zeta^{eff})^2$ decreases with respect to the magnetic fields~\cite{zhang2015magnetic}. Once it becomes negative, unstable modes are present with their energies such as $E^2(k=0)=(m_\zeta^{eff})^2<0$.
In the case of a Mexican-hat-shape potential of a scalar field, as well known, it is excitation of unstable modes around the original minimum to lead to the presence of new stable vacuum.  
Similar excitations of the charged Higgs modes are expected to arise (see the next subsection for details).

At the leading order of $eB$, the squared mass was expressed as~\cite{zhang2015magnetic}
\begin{equation}
\label{eq:magneticmass}
(m_\zeta^{eff})^2 \simeq m_\zeta^2 - v_\perp^2eB.
\end{equation}
where $v_\perp^2$ is the transverse velocity of Higgs fields.
Based on Eq.~(\ref{eq:magneticmass}), a critical magnetic field $B_0$ can be obtained from the condition $(m_\zeta^{eff})^2 \rightarrow 0$, which signals the transition to a less-symmetric
MCLF phase. We will show in Sec.~\ref{sssec:2} that $B_0$ is the right threshold field for the emergence of MCFL.
For instance, by taking $v_\perp^2=1/3$ and adopting $\beta_2$ as
the $\mathscr{O}(1)$ coefficient, we yield the threshold value $eB_0 = 12\beta_2 v^2 \simeq 12v^2$.
This result is in consistent with that achieved from the effective Lagrangian for NG
modes~\cite{ferrer2007magnetic}. Physically, the agreement of two different analyses is due to
the fact that the NG modes correspond one to one to the Higgs modes within the GL framework.
Furthermore, if choosing the CFL gap $v = 50$~MeV, the threshold magnetic field could reach the
order of $10^{18}\text{G}$. Numerically, such a magnitude is comparable to the estimated value of
magnetic fields in the core of neutron stars, but it is still less than the theoretical upper
limit $10^{20}\text{G}$ inside self-bound magnetars~\cite{dong2001,lai1991cold}.
\subsection{Magnetic-induced condensate }
\label{sssec:2}

In the most-symmetric CFL phase, only the diagonal elements are involved in the diquark matrix, namely
\begin{equation}
  \label{eq:phi}
  \Phi =
  \begin{pmatrix}
    d & 0 & 0 \\
    0 & d & 0 \\
    0 & 0 & d
    \end{pmatrix}.\end{equation}
As the color-flavor-locked condensate, $d$ is rotated-charge neutral and the corresponding Higgs
modes are also neutral. Since these modes do not respond to magnetic field directly, the Higgs
masses are still given by $m_\phi$ and $m_\zeta$. Thus, the above-given effective Lagrangian does
not influence the CFL description. Indeed, this point can be verified. By inserting Eq.~(\ref{eq:phi})
into the Lagrangian, there are no essential changes because the traceless potential with
$(m_\zeta^{eff})^2$ is irrelevant to the diagonal matrix element $d$ and its uniform vacuum $v$.

However, it is not the whole story. The ground state of color-flavor-locked-type matter would change
if additional interacting terms and/or condensates are taken into account. The present strategy is
to introduce this condensate. We shall consider the possible condensates except $d$ and study their
roles by an effective Lagrangian with $(m_\zeta^{eff})^2$.
For this purpose, let us firstly investigate the rotated-charge properties of matrix elements in $\Phi$.
According to the definition of the diquark matrix, it might be rewritten as
\begin{equation}
  \label{eq:diquarkmatrix}
  \Phi =
  \begin{pmatrix}
    \Phi_{gb}^{ds} &  \Phi_{gb}^{su} & \Phi_{gb}^{ud} \\
    \Phi_{br}^{ds} &  \Phi_{br}^{su} & \Phi_{br}^{ud} \\
    \Phi_{rg}^{ds} &  \Phi_{rg}^{su} & \Phi_{rg}^{ud}
  \end{pmatrix},
\end{equation}
to account for possible pairings between the quarks with different colors and flavors.
The diagonal elements belong to color-flavor-locked species whereas the non-diagonal elements, as
color-flavor-unlocked species, are not allowed in the ideal CFL phase. Among these non-diagonal
elements, $\Phi_{gb}^{su}$, $\Phi_{gb}^{ud}$, $\Phi_{br}^{ds}$ and $\Phi_{rg}^{ds}$ have the
nonzero charges while $\Phi_{br}^{ud}$ and $\Phi_{rg}^{su}$ are neutral. It can be observed from
the convention for rotated charges of quarks (see e.g., Table~1 in ~\cite{zhang2015magnetic}).
For the charged, non-diagonal elements, they respond to a magnetic field sensitively with
respect to the others. When the magnitude of $B$ is large enough, say $B > B_0$, these elements
are influenced directly and thus unavoidably become excited.

In order to illustrate these excitations more clearly, we return to the familiar language of
the Higgs modes. Keeping in mind the Higgs modes are fluctuations of diquark condensate, our
concerned four elements in Eq.~(\ref{eq:diquarkmatrix}) correspond to the charged Higgs modes
$\zeta^+$ and $\zeta^-$, respectively.
In terms of Higgs octet fields $\zeta^a$, they can be expressed as
\begin{equation}
  \label{eq:diquarkmatrix2}
  \begin{pmatrix}
            &    \zeta^-  & \zeta^-\\
    \zeta^+ &            &      \\
    \zeta^+ &            &
    \end{pmatrix}
=
 \begin{pmatrix}
 & \zeta^1 -i\zeta^2 & \zeta^4 -i\zeta^5 \\
    \zeta^1 + i\zeta^2 &                  &                \\
    \zeta^4 + i\zeta^5   &                 &

     \end{pmatrix}.
\end{equation}
Eq.~(\ref{eq:diquarkmatrix2}) tells us that $\zeta^+$ and $\zeta^-$ are in fact nothing but the
recombination of the Higgs modes around the CFL vacuum. Even though $\zeta^a$ ($a=1, 2, 4, 5$)
do not respond to magnetic background directly, their combinations $\zeta^+$ or $\zeta^-$ does.
As long as the magnitude of $B$ is large, it is safe to regard these magnetic-induced excitations
as the condensates.

For the sake of certainty, we further assume that the condensates from $\zeta^+$ and $\zeta^-$ are
treated as an unique condensate $\delta$. This is a simple but appropriate ansatz since total neutrality
of rotated charge is guaranteed for the color-flavor-locked-type matter.
Instead of Eq.~(\ref{eq:phi}), a more general matrix reads
\begin{equation}
  \label{eq:diquarkmatrix0}
\Phi = \begin{pmatrix} d & \delta & \delta \\ \delta &  d & 0 \\ \delta & 0 & d\end{pmatrix},
\end{equation}
and its VEV may be described by
\begin{equation}
\label{eq:mcflgap}
\text{VEV}(\Phi) = \begin{pmatrix} v & v_\delta & v_\delta \\ v_\delta &  v & 0 \\ v_\delta & 0 & v \end{pmatrix},
\end{equation}
by assigning the vacuum $v_\delta$ to $\delta$.
In the following discussions we shall refer to the less-symmetric ground state as the
\emph{MCFL} phase.

To this stage we start to answer the key question, how the condensate $\delta$ induced dynamically.
After assigning the known vacuum $v$ to $d$, we consider Eq.~(\ref{eq:diquarkmatrix0}) in an effective
Lagrangian with magnetic-dependent coefficient. As the result,
the potential for $\delta$ can be given by
\begin{equation}
\label{eq:deltapotential1}
\mathcal{V}_\delta \sim \alpha' \delta^2 + \beta'\delta^4 + \cdots,
\end{equation}
with
\begin{eqnarray}
  \label{eq:coefficients}
\alpha' \equiv (m_\zeta^{eff})^2,
\end{eqnarray}
and
\begin{eqnarray}
  \label{eq:coefficients2}\beta' \equiv \frac{(m_\zeta^{eff})^2}{6v^2} + \frac{4 m_\phi^2}{3v^2}.
\end{eqnarray}

As the $\delta^2$ coefficient, $\alpha'$ might disappear and even take a negative value in the
presence of magnetic fields.
In the GL analysis with a given order parameter, the sign of coefficient in quadratic term
determines whether or not the order parameter acquire its nontrivial vacuum.
Obviously, the vanishing $\alpha'$ is a signature of the
$\delta$ condensate. Thus, the quantity $B_0$ obtained from $\alpha' \rightarrow 0$ is just the
threshold field for the emergence of MCFL.
For the magnetic field with $B > B_0$, moreover, $\alpha'$ (being negative) is no longer the usual squared mass defined in the known CFL phase. Instead, it reflects the fact that the charged Higgs modes behave as unstable modes
with the negative energy. The unstable modes are excited and eventually leads to a new vacuum and a less symmetric ground state. 
In this sense, the negative coefficient is responsible for the excitation of unstable modes and then the additional condensate $\delta$. 
As for the coefficient $\beta'$, it is partially magnetic dependent. Throughout the current work we
shall consider the situation with positive $\beta'$, otherwise, the existence of $\delta$ is not
theoretically controllable. The possibility of $\beta'< 0$ happens for relatively strong fields, say,
$B \geq 9 B_0$ from numerical estimates, which is beyond the scope of our consideration.

By minimizing the relevant part of Eq.~(\ref{eq:deltapotential1}), the value of $v_\delta$ is given by
\begin{equation}
\label{eq:vacuumexpectation}
v_\delta^2 = \frac{- \alpha'} {2\beta'}.
\end{equation}
such that the potential can be rewritten as a Mexican-hat-shape form
\begin{equation}
\label{eq:deltapotential}
\mathcal{V}_\delta \sim - \alpha'(\delta^2 - v_\delta^2)^2.
\end{equation}
Eq.~(\ref{eq:vacuumexpectation}) is analogous to the form $\mathcal{V}_d \sim - \alpha (d^2 - v^2)^2$
for the $d$ condensate.
As CFL with $d$ appears provided $\alpha$ is negative, the similar dynamical mechanism happens for
the new ground state. It is that the MCFL phase including $\delta$ emerges provided $\alpha$ and
$\alpha'$ are negative.

\subsection{MCFL gap behavior }
\label{sssec:3}

As a simple application, let us examine how the MCFL gap vary with magnetic fields.
Different from the CFL vacuum Eq.~(\ref{cflground}), Eq.~(\ref{eq:mcflgap}) involves both vacuum
expectations simultaneously. Unfortunately, neither $v_\delta$ nor $v$ can be considered as the
MCFL gap directly. As stressed at the beginning of Sec.~\ref{sssec:2}, the present discussion is
limited in the ideal situation with an uniform color-superconducting gap. In this sense, it is
necessary to convert Eq.~(\ref{eq:mcflgap}) into a more proper form for studies of gap behavior.

Without loss of generality, the transformation such as $\Phi \rightarrow \Phi'= U ^\dagger\Phi U$
is feasible with an action $U$ built in the space of diquark with color and flavor.
For our purpose, the ground state with non-diagonal matrix elements needs to be converted into
the state with
\begin{equation}
\label{eq:effectivegap}
\text{VEV}(\Phi') = \begin{pmatrix} v_{eff} & 0 & 0 \\ 0 & v_{eff} & 0 \\ 0 & 0 & v_{eff} \end{pmatrix}.
\end{equation}
Now the diagonal element $v_{eff}$, as a uniform vacuum expectation of diquark condensate, may be
understood as the effective gap, even though it does not correspond to the MCFL gaps calculated in
NJL model in a very strict sense. Due to $U \in SU(3)_C \times SU(3)_F$, a determinant manipulation
for two matrices yields the relationship like
\begin{equation}
  \label{eq:det}
\text {det}  \begin{pmatrix} v & v_\delta & v_\delta \\ v_\delta &  v & 0 \\ v_\delta & 0 & v \end{pmatrix}=  \text {det} \begin{pmatrix} v_{eff} & 0 & 0 \\ 0 & v_{eff} & 0 \\ 0 & 0 & v_{eff} \end{pmatrix}.
\end{equation}
In such a simple way, the gap behavior in the MCFL phase may be reproduced qualitatively.

\begin{figure}[ht]
\centering
\includegraphics[width=0.7\textwidth]{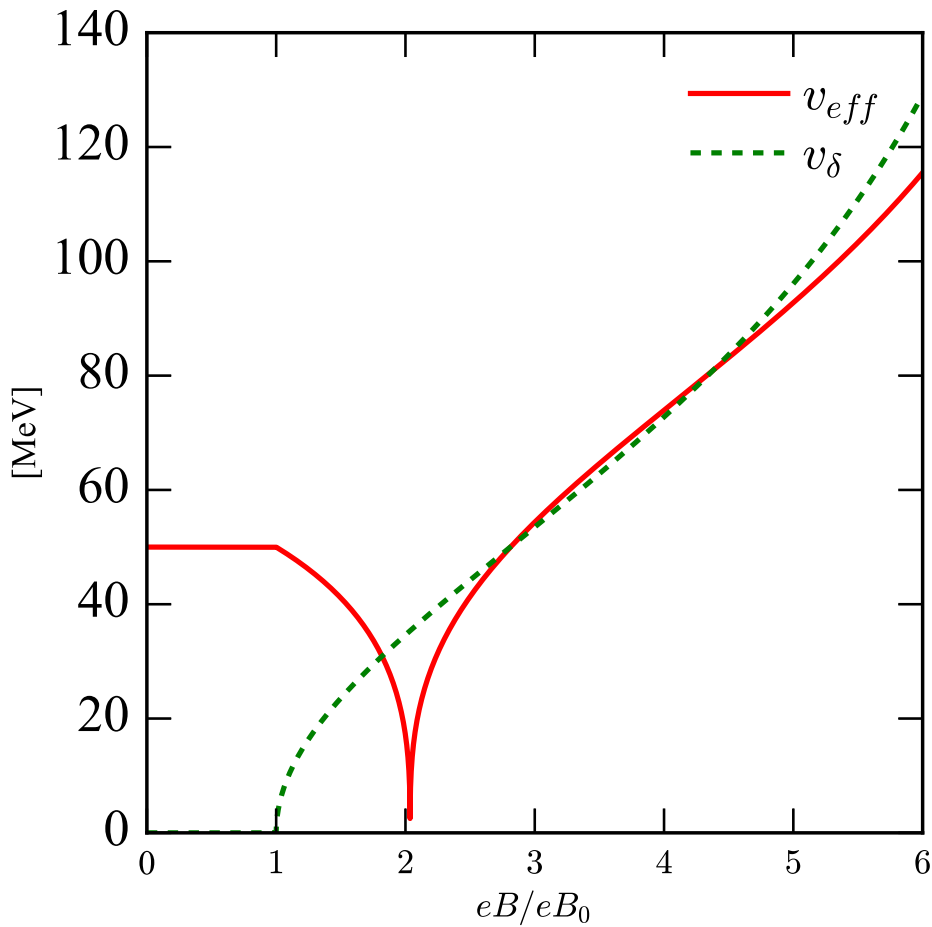}

\caption{Schematic figure of the effective gap (red, solid) and the expectation of magnetic-induced condensate (green, dashed) as functions of magnetic field.}
	\label{fig:1}
\end{figure}

Within the present framework, $v$ has been regarded as a known parameter, or the CFL gap.
By using the simplification $\beta_1=\beta_2=\beta = 1$ and the numerical estimate Eq.~(\ref{eq:magneticmass}),
the vacuum defined by Eq.~(\ref{eq:vacuumexpectation}) is expressed as
\begin{equation}\label{eq:vdelta2}
v_\delta = v \Big(\frac{3eB-36v^2}{108v^2-eB}\Big)^{1/2}.
\end{equation}
Eq.~(\ref{eq:vdelta2}) is valid only for $B > B_0$ and $B < 9B_0$. Numerically, the former
corresponds to $eB > 12v^2$, which can be seen from the numerator, and the latter to $eB < 108v^2$
from the denominator in Eq.~(\ref{eq:vdelta2}). Again, as mentioned before, the latter originates
from the constraint of $\beta'> 0$.
As shown by the dashed line in Fig.~\ref{fig:1}, $v_\delta$ keeps an increasing function of
magnetic field.

For the value of $v_{eff}$, it is required to be positive (as an effective gap).
Depending on varying magnetic fields, there are the different results derived from Eq.~(\ref{eq:det}).
For weak magnetic fields (but larger than the threshold value), the magnitude of $v_\delta$ remains
small relative to $v$. In this situation, Eq.~(\ref{eq:det}) is reduced to
\begin{equation}
\label{eq:veff1}
v_{eff} =  (v^3-2 vv_\delta^2)^{1/3},
\end{equation}
so that the effective gap is firstly suppressed with introducing $eB$ (see Fig.~\ref{fig:1}).

With increasing magnetic fields, on the other hand, the magnitude of $v_\delta$ is possible to
become dominant (relative to $v$). In this situation, we have to perform the absolute value of
Eq.~(\ref{eq:veff1}) by hand. By inserting the result of $v_\delta$, we obtain the $B$ dependence
\begin{equation}
\label{eq:veffchanged}
v_{eff} = v \Big(\frac{7eB-180v^2}{108v^2-eB}\Big)^{1/3}.
\end{equation}
From the numerator of Eq.~(\ref{eq:veffchanged}), the vanishing gap is found to be $eB\simeq 180v^2/7$
($B \simeq 2.1 B_0$).
When the magnetic field exceeds this value, as shown in Fig.~\ref{fig:1}, a steadily increasing function
is observed for the effective gap. Thus, this inflection point may reflect the ``transition" from a weak
decreasing tendency to a strong increasing tendency. Of course, such an artificial point originates from
uncertainties behind our estimate hence its result should not be taken too seriously.

To this end, we would like to address on the magnetic-field dependence of $v_{eff}$. In the
present GL analysis, the so-called ``strong field" means that $B$ is in a regime
$2.1 B_0 < B \ll 9 B_0$, as shown in Fig.~\ref{fig:1}. Note that, for such a finite regime,
the ratio $v^2/eB$ can be safely regarded as a small quantity and the numerator of
Eq.~(\ref{eq:veffchanged}) provides a more obvious dependence than the denominator part.
Roughly, we might expand the numerator by the power series of $v^2/eB$. With respect to the
maximum value, say, $\sim 100$ MeV, possible for color-superconducting gap, the dimensionless
form of $v_{eff}$ is simplified as
\begin{equation}\label{eq:expansion1}
 v_{eff} \sim 1 - a_1 \frac{v^2}{eB} + \cdots \ ,
\end{equation}
at the leading order. There, the value of $a_1$ is found to be positive which ensures the
monotonic tendency of effective gap.

On the NJL-model side, the equation of MCFL gap(s) had been derived analytically for strong
magnetic fields, where only the lowest Landau level is occupied.
The ``strong field" means that $eB$ is approximately the order of the square of quark chemical
potential $\mu_q^2$. Particularly for the $\widetilde{Q}$-relevant gap, it was discovered that
the dimensionless form like
$\Delta \sim \text{exp}{[-{3\pi^2 \Lambda^2}/{(g^2\mu_q^2+ g^2 eB)}]}$~\cite{ferrer2006color},
where $g$ stands for a dimensionless coupling coefficient and $\Lambda$ is a cutoff of energy.
For the above result, we should devote ourselves to the $eB$ dependence rather than the $\mu_q$ dependence.
In order to ``remove" the $\mu_q$ dependence formally, the limit of $\mu_q^2 = eB/2$
needs to be adopted which was proved important for the self-consistency in our concerned situation with a uniform
gap~\cite{sen2015anisotropic}. By considering the specific limit and employing the power-series method again, the
leading-order expansion of $\Delta$ is formally written as $\sim 1 - a_2 \Lambda^2/eB + \cdots$.
Since $\Lambda^2/eB$ itself is not small enough, the value of $a_2$ might be significantly deviated from $a_1$ in
Eq.~(\ref{eq:expansion1}). Even so, $a_2$ is still required to be positive. In this sense, the dependence of
$\Delta$ on strong magnetic fields displays a qualitative similarity as that of $v_{eff}$.

The similarity becomes invalid completely in vicinity of the inflection point shown in Fig.~\ref{fig:1}. 
In the NJL numerical calculations the de Haas-van Alphen oscillations of MCFL gap(s) were
observed~\cite{ferrer2005magnetic,fukushima2008color}, although some recent studies suggest that parts of
unphysical oscillations may be eliminated by the appropriate regularization scheme ~\cite{allen2015magnetized}.
Obviously, the exotic gap behavior failed to be reproduced in our result. 
In fact, the present-concerned MCFL phase does not strictly correspond to that studied by using phenomenological quark models. There are some of differences between our model-independent treatment and the usual
NJL calculations.
Among them the essential one is that no Landau Level for quarks is introduced in the present analysis, which might explain why we are not able to yield gap oscillations in Fig.~\ref{fig:1}. 
We are actually concerned the magnetic response of charged Higgs modes, rather than that of charged quarks. 
Of course, the GL treatment is a prior analysis based on symmetry consideration purely. It only allows for qualitative discussions since quark dof are not incorporated explicitly. 
In addition, the magnetic dependence of $\alpha'$ is given at the leading order of $eB$, as shown in Eq.~(\ref{eq:magneticmass}). 
It should be no longer valid when $eB$ is large enough, say, the magnitude with order of $\mu_q^2$.
Superficially one might overcome this shortcoming by extending Eq.~(\ref{eq:magneticmass}) to a more general expansion such as $\alpha'=(m_\zeta^{eff})^2 = m_\zeta^2 - v_\perp^2eB + C_2 (eB)^2 + \cdots$. Nevertheless, it is nontrivial to handle the expansion for the strong fields.  
Within the GL framework, not only the coefficients like $C_2$ but also the transverse velocity $v_\perp$ are still missing in the strong-field limit. The NJL result for $v_\perp$ in this limit is clearly deviated from $1/\sqrt{3}$~\cite{sen2015anisotropic}.
Therefore, we are not able to describe detailed properties of the homogenous MCFL phase quantitatively.

\section{Magnetic-induced vortex formations}
\label{sec:3}

By introducing the magnetic-induced condensate, we have defined the less-symmetric ground state and
studied behavior of the effective gap in a homogenous MCFL phase. Before exploring possible topological
vortices consisting of MCFL, we give a brief review on the known vortex solutions consisting of CFL.

As seen in Eq.~(\ref{cfl}), the original QCD symmetry $G$ is broken to
the color-flavor locked symmetry $H$.
For the CFL phase, the diquark matrix $\Phi$ can be parameterized in the
topological space
\begin{equation}
  \label{eq:cflvortexgroup}
  \frac{G}{H} \simeq \frac{SU(3) \times U(1)_B}{Z_3}  \simeq U(3),
\end{equation}
where $Z_3$ is a discrete symmetry.
Since the symmetry $U(1)_B$ is broken spontaneously, on the one hand, an usual superfluid vortex is generated.
In the cylindrical coordinates, the spatial configuration for a minimal-wound CFL vortex reads
\begin{equation}
  \label{eq:bvortexphi}
\Phi =vf(r)e^{i \theta} \texttt{diag}(1,1,1),
\end{equation}
and equivalently
\begin{equation}
 d = vf(r)e^{i\theta}, \label{eq:bvortex}\end{equation}
due to Eq.~(\ref{eq:phi}). There, the polar angle $\theta$ originates from $U(1)_B$ breaking and
$f(r)$ is the profile function with the boundary
conditions $f(0) = 0$ and $f(\infty) =1$.

On the other hand, a particularly-interesting topological object has been suggested for
CFL more recently. Because of the non-Abelian structure of Eq.~(\ref{eq:cflvortexgroup}), the
diagonal elements of $\Phi$ are no longer degenerated in the resulting object and are called as
the ``non-Abelian'' vortices. For instance, the typical minimal-wound solution is given
as~\cite{nakano2008non,eto2009color}:
\begin{equation}
  \label{eq:nvortex}
  \Phi = v\begin{pmatrix}
   f(r)e^{i\theta} & & \\ & g(r) & \\ & & g(r)
  \end{pmatrix},
\end{equation}
which may be described by two of independent vortex solutions also~\cite{balachandran2006semisuperfluid}.
An important property of such kind of CFL vortices
is that the color-flavor-locked symmetry breaking pattern might change. In the vicinity of core of CFL
vortices, it has been pointed out that the locked symmetry $H={SU(3)_{C+F}}$ is broken to
$H' =U(1)_{C+F} \times SU(2)_{C+F}$ \cite{nakano2008non,vinci2012spontaneous}.
The NG modes associated with this breaking are of the orientational zero modes and they appear in the
topological space
\begin{equation}
  \label{eq:cp2}
 \frac{H}{H'} = \frac{SU(3)_{C+F}}{U(1)_{C+F} \times SU(2)_{C+F}} = CP^2.
\end{equation}

At this moment, we want to emphasize that the formation of CFL vortices itself bores no relation to an
external magnetic field. In other words, both $U(1)_{B}$ superfluid vortex and non-Abelian vortices are
generated spontaneously regardless of magnetic fields.
Once CFL has been replaced by MCFL, it is necessary to reexamine the formation mechanism of vortices in
the presence of an applied field $B$, as we shall show in the following.

\subsection{Superfluid vortex formation }
\label{sssec:4}

Assuming that MCFL emerges inside the core of CFL non-Abelian vortices, the remaining subgroup $H'$ is
possible to be further broken to $M =SU(2)_{C+F}$ by enforcing the magnetic-induced breaking Eq.~(\ref{cfl2}).
In this case, the relevant order parameter for MCFL is parameterized in the topological space
\begin{equation}
  \label{eq:mcfsymm}
 \frac{H'}{M} = \frac{SU(2)_{C+F} \times U(1)_{C+F}}{SU(2)_{C+F}} = U(1)_{C+F},
\end{equation}
rather than the space $H/M$.
Because of the important role played by $\delta$ at large $B$, $U(1)_{C+F}$
breaking could be responsible for the $\delta$ condensate. Consequently, a superfluid vortex (string) for
MCFL is generated from $U(1)_{C+F}$ breaking, just like an usual $U(1)_B$ vortex for CFL. As the simplest
physical picture, this possibility shall be taken as our first step.

\begin{figure}[ht]
\centering
\includegraphics[width=0.7\textwidth]{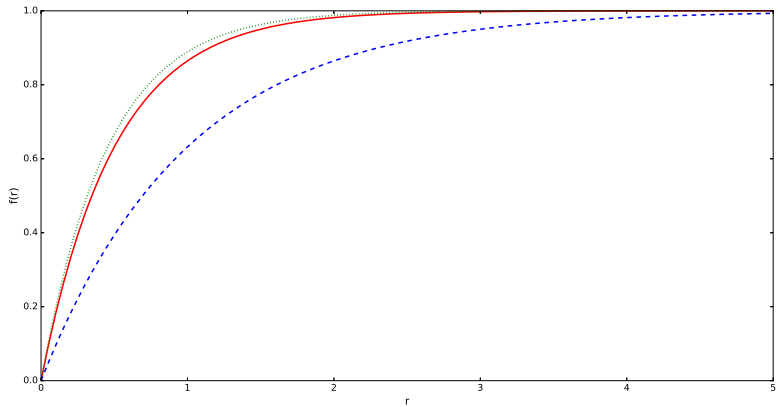}
	\caption{The profile functions of $\delta$ string with $eB = 4eB_0$ (red, solid) and $eB =
    5 eB_0$ (green, dashed) and the profile function of $d$ string (dotted).}
	\label{fig:2}
\end{figure}

Similar as the $d$ string with Eq.~(\ref{eq:bvortex}), the vortex configuration for the resulting $\delta$
string is
\begin{equation}
	\delta= v_\delta f(r) e^{i\theta},\label{mcflvortex}
\end{equation}
where the phase angle $\theta$ arises from $U(1)_{C+F}$ breaking.
For our purpose, it is convenient to employ the reduced Lagrangian of $\delta$
\begin{equation}
\label{eq:mcflvorticehamilton}
 \mathcal{L}_\delta = (\partial \delta)^+ (\partial \delta) - \frac{\alpha'}{4}\delta^2 - \frac{\beta'}{4}\delta^4,
\end{equation}
instead of the original GL formalism with the matrix $\Phi$.
By inserting Eq.(\ref{mcflvortex}) into Euler-Lagrange equation, we obtain the profile function from
\begin{equation}
\label{eq:profilefunction}
 f'' + \frac{f'}{r} -\frac{f}{r^2} - (\frac{\alpha'}{4} + \frac{\beta'}{2} v_\delta^2 f^2)f=0,
\end{equation}
where $f'$ and $f''$ denote the first- and the second-order derivatives of $f(r)$ with respect to $r$,
respectively. Eq.~(\ref{eq:profilefunction}) is magnetic-field dependent as the coefficients $\alpha'$
and $\beta'$ are involved.
In Fig.~\ref{fig:2}, the profiles of $\delta$ string with two typical magnetic-field values (see the
solid line and the dashed line) are plotted. With varying magnetic fields, the shape of the profiles
is changed.

For a comparison with the CFL case, we consider a simple $U(1)_B$ vortex solution only and yield the
profile equation for the $d$ string
\begin{equation}
  \label{eq:bvortexprofile}
  f'' + \frac{f'}{r} -\frac{f}{r^2} - (\frac{\alpha}{3} + \frac{8}{3}\beta v^2 f^2)f=0,
\end{equation}
from Eq.(\ref{eq:bvortex}) in a similar way.
The corresponding profile function, being magnetic-field independent, is shown by the dotted line in
Fig.~\ref{fig:2}. Now, a significant shape difference between the $d$- and $\delta$-profiles is observed.
Utilizing the characteristic radii of normal core in the two kinds of $U(1)$ vortices, it is clear that
the characteristic radius $R_d$ is far larger than the radius $R_\delta$.
This result is not surprising. According to our assumptions, the $\delta$-string generation happens in
the core region of non-Abelian vortices. It means that the vortex-core size of the latter is required
to be larger than that of $\delta$ string. In the scope of CFL vortex solutions, the radius of $U(1)_B$ vortex, $R_d$, exceeds the mean radius of non-Abelian vortices. In this sense, the
above requirement is turned into a simple relation, $R_d > R_\delta$. We show this in Fig.~\ref{fig:2}.

Moreover, it is possible to derive the formation condition for $\delta$ string in a more formal way.
It is well known that the magnitude of $R$, as a correlation length of the concerned condensate, may be
estimated by an inverse mass of the Higgs modes associated with $U(1)$ breaking~\cite{vilenkin2000cosmic}.
In the CFL case, the inverse mass of $\phi$ mode determines $R_d$ essentially.
By considering the Lagrangian of $d$ (see the Eq.~(\ref{eq:vortonb}) below), $R_d$ is estimated to be
$(-\alpha/3)^{-1/2}$.
Similarly, $R_\delta$ is found to be about $(-\alpha'/4)^{-1/2}$ from Eq.~(\ref{eq:mcflvorticehamilton}).
Based on these estimates, the relation $R_d > R_\delta$ is further turned into
$- 3\alpha' > - 4\alpha$. It behaves as the forming condition for $\delta$ string and corresponds to the region of magnetic field 
$eB > - 4\alpha+12\beta_2 v^2$. Numerically, the region is about $eB > \frac{7}{3} eB_0$ with the simplification $\beta_1=\beta_2=\beta = 1$.
The situations $eB = 4eB_0 $ and $5eB_0$ given in Fig.~\ref{fig:2} clearly fulfill the condition.
Also, it is time to explain the shape change of $\delta$-string profiles. Utilizing the language of
characteristic radius again, this result is easily understood from the $B$ dependence of $R_\delta$.
With increasing magnetic field, the value of $-\alpha'$ becomes larger such that the value of
$R_\delta$ is small relatively.
This is the reason why the spatial size of $\delta$-string profile tends to be suppressed as shown
in Fig.~\ref{fig:2}.

Finally, we study the kinetic energy of $\delta$ string per length unit. Physically, such kind of
linear tension might be roughly given by the area of profile function versus $r$. From Fig.~\ref{fig:2},
the difference in areas under two $\delta$-profiles indicates the existence of the $B$ dependence of
string tension.
Without loss of generality, the definition of string tension is expressed as
\begin{equation}
\mathcal{T} = \int^{2\pi}_{0}d\theta \int^L_{R_\delta}
\mathcal{H} rdr
\label{eq:tension},
\end{equation}
where $\mathcal{H}$ denotes the system Hamiltonian.
Beside the vortex-core radius $R_\delta$, $L$ is introduced to account for the total radius of a
superfluid vortex. In this subsection, it is treated as a cut-off constant at which the boundary
conditions are $f =1$ and $f' =0$.
By using the $\delta$-profile equation Eq.~(\ref{eq:profilefunction}), we ignore the constant
contribution and give the asymptotic expression of string tension
\begin{equation}
  \label{eq:tension1}
  \mathcal{T} \sim v_\delta^2 ln\frac{L}{R_\delta},
\end{equation}
which can also be derived from the quantization of $U(1)$ string.

Based on Eq.~(\ref{eq:tension1}), let us examine the magnetic-field dependence of $\delta$-string tension.
For a stronger field, on the one hand, the vacuum expectation $v_\delta$ increases steadily.
This point is easily found in Eq.~(\ref{eq:vacuumexpectation}) and/or Fig.~\ref{fig:1} in Sec.~\ref{sec:2}.
On the other hand, the vortex-core radius $R_\delta$ has an opposite dependence, i.e. its value slightly
decreases as shown in Fig.~\ref{fig:2}). With increasing $B$, therefore, the logarithmic-divergent
tendency becomes much more obvious for the tension energy. 

\subsection{Formation of vorton structure }
\label{sssec:5}

Now we turn to a more complicated situation where both $\delta$- and $d$-condensate exhibit the
spatial-dependent properties simultaneously. Instead of the above-discussed picture, there exists a
theoretical possibility that the two strings from $U(1)$ breaking allow for the existence of vortons,
a topologically and energetically stable object. The scenario of vortons was first considered in the
scope of cosmic
string~\cite{vilenkin2000cosmic,witten1985superconducting,davis1988physics1,davis1988physics2,haws1988superconducting}.
For color superconducting matter of dense QCD, the vorton formation was studied in the case of introducing the
condensates of two NG modes, say, $K^0$ and $K^+$, in the CFL environment. As the $K^+$ and $K^0$ condensed strings are generated from $U(1)$ breaking, their coexistence was pointed out to support the stable vortons under somewhat conditions~\cite{kaplan2002charged,buckley2002superconducting}. 
Even though these condensates due to flavor asymmetry are not mentioned, the present-concerned system involves both two condensates as well. As long as the $d$- and
$\delta$-condensate are assumed to arise from $U(1)_B$ and $U(1)_{C+F}$ breaking respectively, the essential physics
behind magnetic color-flavor-locked matter should share some analogy with that discussed in the literature.

Our starting point is that the $d$-string is generated while the $\delta$ condensate is expected to emerge
at the $d$-string centre. The former can be regarded as a usual, straight vortex temporarily parallel to the
$z$ direction. Similar to the Eq.~(\ref{eq:bvortex}), it has the form of $d = d(r)e^{i\theta}$.
Along the $z$ direction, the $\delta$ condensed field carry non-vanishing charge and current as its emergence
is inside the string core. Without losing generality, the ansatz for such a nontrivial $\delta$ vortex solution reads
\begin{equation}
  \label{eq:delta}
  \delta =  e^{i(kz+\omega t)}\delta(r),
\end{equation}
which is different from the $\delta$ string in Sec.~\ref{sssec:4}. The frequency $\omega$ contributes to the conserved
Noether charge $Q$ via
 \begin{equation}
  \label{eq:vortonquantumq}
   Q = \omega\int dz \int dS \delta^2,
 \end{equation}
where $S$ denotes the area being perpendicular to $z$ axis. Similarly, the wave number $k$ will contribute to the
current $J$ along $z$ axis via $J =k\int dz \int dS \delta^2$ in the minimal-wound case.

We first investigate the necessary and sufficient conditions for existence of $\delta$ condensate
at the $d$-string centre.
For this purpose, we will consider the simplest theory with two order parameters, say
a $U(1) \times U(1)$ model Lagrangian $\mathcal{L}(d,\delta)= \mathcal{L}_d +
\mathcal{L}_\delta + \mathcal{L}_{d\delta}$.
The Lagrangian of $d$ field is easily written in the Mexican-hat form
\begin{equation}
  \label{eq:vortonb}
  \mathcal{L}_d  = (\partial d)^+ (\partial d) +\frac{\alpha}{6 v^2}(d^2 - v^2)^2,
\end{equation}
where the known vacuum Eq.~(\ref{eq:dvaccum}) has been used to eliminate the coefficient $\beta$.
For the $\delta$ field, as stressed in Sec.~\ref{sec:2}, it is totally decided by the coefficient
$\alpha'$ rather than $\alpha$. The Lagrangian has been given by Eq.~(\ref{eq:deltapotential1}), equivalently
\begin{equation}
  \label{eq:vortond}
  \mathcal{L}_\delta = (\partial \delta)^+ (\partial \delta) + \frac{\alpha'}{8 v_\delta^2}(\delta^2 - v_\delta^2)^2.
\end{equation}
The term with $d$ and $\delta$ mixing, being important for a $U(1) \times U(1)$ model,  may be formally written as
\begin{equation}
  \label{eq:vortoninter}
  \mathcal{L}_{d\delta} = -\lambda d^2 \delta^2,
\end{equation}
where $\lambda$ is required positive.
Recalling the original GL Lagrangian with the matrix $\Phi$, this term needs to
be associated with the contributions from $\alpha$ and $\beta_2$ (exactly, $\alpha'$) at the same time. In order to guarantee
that the resulting Lagrangian is theoretically controllable and resembles a
simple $U(1) \times U(1)$ model, we pick upon the relevant terms
in expansion of Eq.~(\ref{gl}) and define the coefficient as
\begin{equation}
  \label{eq:vortoninter1}
  \lambda = -\frac{1}{3 v^2}(\alpha' +\frac{\alpha}{3}).
\end{equation}

Apparently, introducing the ansatz in Eq.~\ref{eq:delta} leads to the changes in the formalism.
In particular, the effective potential for $\delta$ field becomes
\begin{equation}
  \label{eq:vortond1}
  \mathcal{V}_\delta= -\frac{\alpha'}{8v_\delta^2} [\delta^2 - (v_\delta^2 - \frac{8v_\delta^2}{\alpha'}(\omega^2 -k^2))]^2 \ ,
\end{equation}
rather than the formalism Eq.~(\ref{eq:deltapotential1}).
In the vacuum with $d \neq 0$ and $\delta = 0$, the effective potential provides the additional
contribution to the Lagrangian $\mathcal{L}_d$. To guarantee the $d$-field symmetry is broken such
that $d \neq 0$, it is necessary to require that the vacuum contribution (constant term) is positive.
By considering Eqs.~(\ref{eq:vortonb}) and (\ref{eq:vortond1}), the condition reads
\begin{equation}
\label{eq:vortonb2}
  \frac{\alpha}{6} v^2 < \frac{\alpha'}{8v_\delta^2}(v_\delta^2  - \frac{8v_\delta^2}{\alpha'}(\omega^2 -k^2))^2.
\end{equation}
On the other hand, the Lagrangian of $\delta$ needs to be examined in such a vacuum. To guarantee the
$\delta$-field symmetry remains unbroken such that $\delta = 0$, it is required that the quadratic
coefficient of $\delta$ is positive. By using Eq.~(\ref{eq:vortond1}) again and considering the mixed
term, we yield the another condition such as
\begin{equation}
  \label{eq:vortonb3}
  \lambda v^2 + \frac{\alpha'}{4}- \omega^2 +k^2 > 0.
\end{equation}

The conditions Eqs.~(\ref{eq:vortonb2}) and (\ref{eq:vortonb3}) are not sufficient to yield the vacuum
with $d = 0$ and $\delta \neq 0$ which emerges at the centre of $d$ string. Contrary to the
Eq.~(\ref{eq:vortonb3}), the requirement that the $\delta^2$ coefficient is negative seems to result in
\begin{equation}
  \label{eq:vortonb4}
\omega^2 - k^2 -\frac{\alpha'}{4} > 0 \ .
\end{equation}
Here, the mixed term with $\lambda$ does not take effect at classical level. Note that $\delta$ is of
the quantum solution essentially, its temporal gradient energy cost should also be taken into account.
For the ground state of $\delta$ condensate, one may add such a perturbation term as $e^{i\nu t}$ and
substitute it into Eq.~(\ref{eq:vortond1}). Note that the mode associated with $\nu$ is actually required
to possess a negative eigenvalue~\cite{vilenkin2000cosmic,haws1988superconducting}, a more accurate form
of the sufficient condition is given by
\begin{equation}
  \label{eq:vortonb5}
  \omega^2 - k^2 -\frac{\alpha'}{4} > \sqrt{- \frac{2}{3}\alpha \lambda v^2}.
\end{equation}
Only if the conditions Eqs.~(\ref{eq:vortonb2}), (\ref{eq:vortonb3}) and (\ref{eq:vortonb5}) are
satisfied at the same time, there exist not merely the $d$ condensate in the region with large $r$, but
also the $\delta$ condensate in vicinity of the region with $r \rightarrow 0$.

Then, we derive the profile equations based on the Lagrangian $\mathcal{L}(d,\delta)$.
It is easy to obtain the profile function of $d$ string from
\begin{equation}
  \label{eq:deuler}
  d'' +\frac{d'}{r} - \frac{d}{r^2} - (\lambda \delta^2 + \frac{\alpha}{3})d + \frac{\alpha}{3v^2}d^3 = 0,
\end{equation}
where the mixed term has played its role.
For the profile function of $\delta$ condensate, its motion equation may be written as
\begin{equation}
  \label{eq:beuler}
  \delta'' +\frac{\delta'}{r} - \frac{\delta}{r^2} - (k^2 - \omega^2)\delta - (\lambda d^2 + \frac{\alpha'}{4})\delta + \frac{\alpha'}{4v_\delta^2}\delta^3 = 0 \ .
\end{equation}
Comparing with the $\delta$-string result Eq.~(\ref{eq:profilefunction}) obtained in Sec.~\ref{sssec:4},
not only the mixed term, but also the contributions from $k^2$ and $\omega^2$ are involved.
In principle, both profiles would be affected by an external magnetic field. For the $d$ string, it comes
from the coefficient $\lambda$. Even so, it is difficult to investigate the magnetic-field dependence
because of the severe conditions Eqs.~(\ref{eq:vortonb2}), (\ref{eq:vortonb3}) and (\ref{eq:vortonb5}).
For certainty, let us consider the limit of $\omega^2 = k^2$ which corresponds to a
critical situation for our concerned $\delta$ condensate ~\cite{lemperiere2003behaviour}.
In this case, the magnetic-field values fulfilling three of conditions are found to be ``confined" to a
rather narrow region.
Numerically, the appropriate value of $eB$ is about $2 eB_0$ with the simplifications used above.
When a magnetic field is out of (exactly larger than) it, the vortex solution with Eq.~(\ref{eq:delta})
is likely to decay into an usual string discussed in Sec.~\ref{sssec:4}.

\begin{figure}[ht]
\centering
\includegraphics[width=0.7\textwidth]{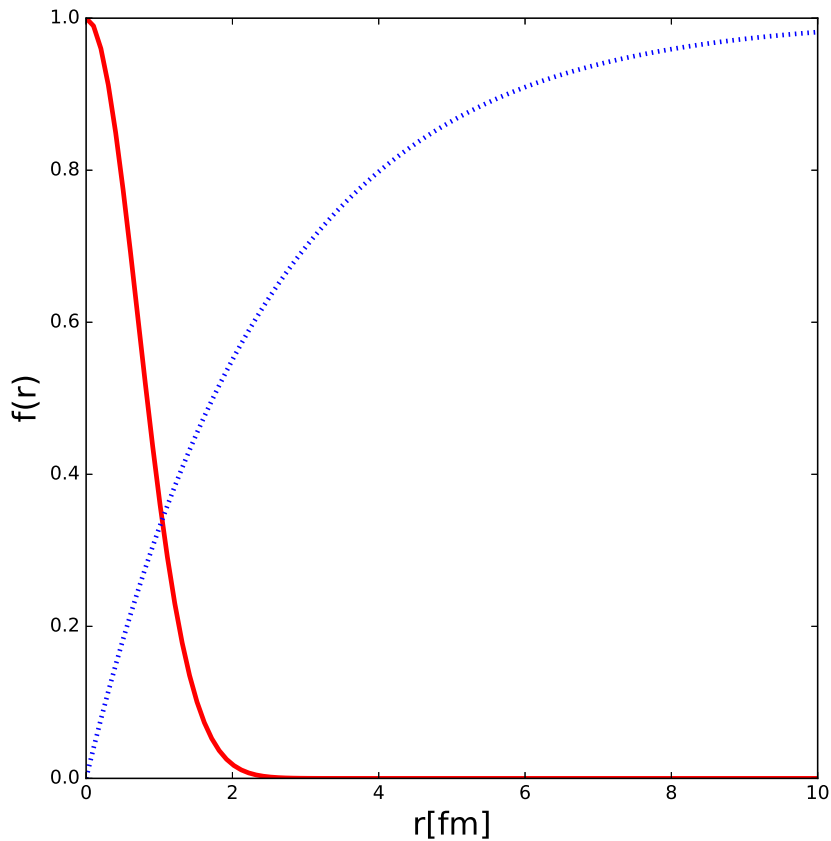}
	\caption{Dimensionless profile functions $f(r)=d(r)/v$ (blue,
    dotted) and $f(r)=\delta(r)/v_\delta$ (red, solid) with $eB= 2 eB_0$ in the $\omega^2 = k^2$ limit.
     }
	\label{fig:3}
\end{figure}

With the boundary conditions $\delta(r \rightarrow 0) = v_\delta$ and $\delta(r \rightarrow \infty) = 0$
as well as the usual boundaries for $d$, we plot the two of profile functions in Fig.~\ref{fig:3}.
It is clear that the non-vanishing $\delta$ fields emerge only inside the core region of $d$ string.
The similar behavior had been obtained in the literature and it opens the possibility of a nontrivial
vorton structure. Also, Fig.~\ref{fig:3} is worthy of discussions by utilizing the language of the
total radius $L$.
From a comparison of the spatial thickness, it is observed that $L$ for $d$ string is far larger than
$L_\delta$ for $\delta$ string. This result implies that, when the vorton structure is introduced, no
complicated curvature effect appears so that the energy of the system can be calculated in a relatively
simple manner.

Finally, it is time to introduce a proper spatial configuration and construct a vorton state.
To obtain a finite energy for the $d$ vortex, the simple choices are that it exists in a finite
container or it forms a closed circular loop.
Our concerned spatial configuration is the latter case. When a straight $d$ string is bent to a
closed loop (ring), $z$ denotes the direction along a ring and the ring radius becomes $L$.
Consequently, the phase change of $\delta$ field and thus the charge/current actually happen in
the ring direction. This is just the picture of vortons (also known as string loops, vortex rings).
In Ref.~\cite{bedaque2011vortons}), a vorton state was illustrated in Fig. 1 for the $K^0$ and
$K^+$ fields. In the present situation, the $d$ field changes in the arrow of $K^0$ while the
$\delta$ field, like $K^+$, does along the direction of ring.

Now that a linkage of two vortices has been realized, the next step is to investigate the energy of a vorton state
and analyze its stability. For the demonstrative purpose, we will discuss the contributions from $d$ and
$\delta$, respectively. For the $d$-string, its length is $2 \pi L$ as a closed loop.
Suppose its linear tension is $\mathcal{T}_d$, the energy for $d$ vortex is given by
$\mathcal{E}_d = 2\pi L \mathcal{T}_d$. As seen this energy is mainly decided by $L$. 
If the system were made up of $d$ singly, thus, it would prefer to shrink
rather than expand.
From the Hamiltonian for $\delta$ field, on the other hand, the $\delta$-vortex energy is expressed as 
\begin{equation}
  \label{eq:deltah}
  \mathcal{E}_\delta = \int dz \int dS  (\nabla_r \delta)^2 + (k^2 + \omega^2)\delta^2 + (\lambda d^2 + \frac{\alpha'}{4})\delta^2 - \frac{\alpha'}{8v_\delta^2}\delta^4.
\end{equation}
By considering the profile equation Eq.~(\ref{eq:beuler}), it is reduced to
\begin{equation}
  \label{eq:energydelta}
  \mathcal{E}_{\delta} =  2\pi L\frac{\alpha' \Sigma_4}{8v_\delta^2} + 4 \pi L \omega^2 \Sigma_2,
\end{equation}
where $\Sigma_2$ and $\Sigma_4$ are short for
$\Sigma_2 = \int dS \delta^2$ and $\Sigma_4 = \int dS \delta^4$ respectively.
In view of the fact that the quantity $Q$ defined by Eq.~(\ref{eq:vortonquantumq}) is
conserved during variation of $L$, the energy can be further simplified as
\begin{equation}
  \label{eq:energydelta1}
  \mathcal{E}_{\delta} =  2\pi L\frac{\alpha' \Sigma_4}{8v_\delta^2} + \frac{Q^2}{\pi L \Sigma_2}.
\end{equation}
Note that the last term in RHS of Eq.~(\ref{eq:energydelta1}) has the
$L^{-1}$ behavior. 
This implies that the system made up of $\delta$ might prefer expanding of a vorton in somewhat situations.

Now, it is the competition between two kind of tendencies, shrinking and expanding, to make an energetically-stable vorton state possible. For the total energy $\mathcal{E} = \mathcal{E}_\delta + \mathcal{E}_d$, we minimize it versus $L$ and obtain the stabilized radius from  
\begin{equation}
\label{eq:vortonr}
 L_0^2 = \frac{Q^2}{2\pi^2\Sigma_2(\mathcal{T}_d +
   \frac{\alpha' \Sigma_4}{8v_\delta^2})}.
\end{equation}
At the radius $L_0$, the stable energy for a vorton state behaves as a
function of $Q$ and $L$, say, $\mathcal{E}_0 = 2 Q^2/(\pi L_0 \Sigma_2)$.
Different from a straight string, the vorton energy is no longer divergent. In other words, the infrared divergence of $\mathcal{T}_d$, c.f. Eq.~(\ref{eq:tension1}), has been eliminated due to the vorton structure.

It is crucial to further examine stable vortons in the astrophysical ``Laboratories" such like the interior of compact stars.
Unfortunately, there still exist some of important aspects which are not dealt with in the present study. 
First of all, the profile behaviors Fig.~\ref{fig:3} is obtained from the critical situation $\omega^2 = k^2$ (but the discussion of vorton energy does not depend on this limit). When $\omega^2 \neq k^2$ is taken into account, existences of $Q$ and $J$ might be important. For instance, a vorton state might be destroyed by a large current $J$ in the context of cosmic string ~\cite{vilenkin2000cosmic}.
Secondly, it is assumed in this section that the $\delta$ field stems out of $U(1)_{C+F}$ breaking.
If the rotated electromagnetic mechanism remains valid, nevertheless, the $\delta$ condensate itself could be classified into two species, say, these from $\zeta^+$ and $\zeta^-$excitations, in the sense of $U(1)_{\widetilde{Q}}$. 
Naively, two kinds of fields should have the opposite orbiting directions in the presence of magnetic fields.
In this case, it is necessary to build more realistic models and study the electromagnetic properties for vortons. 
At the same time, the geometry of vorton structure needs to be reexamined. These details will leave for future studies. We hope that the basic forming mechanism presented here may survive various corrections.

\section{Summary and discussions}
\label{sec:4}

Within the GL framework, we investigated effect of a rotated, external magnetic field on
color-flavor-locked-type matter. Instead of pretending to derive a Lagrangian, we attributed
magnetic effect to the response of charged Higgs modes.
Since the most-symmetric CFL phase is regarded as the known solution, our treatment is of an effective theory
concentrating on corrections to the known results, namely, changes in non-Abelian Higgs spectrum and the
traceless part of GL potential. With help of the magnetic-induced condensate $\delta$, it can demonstrate a
less-symmetric MCFL phase without introducing more details on CFL such like weak-coupling analyses for the
CFL gap.
Moreover, we explored possible formation of vortices consisting of MCFL.
Assuming that such kind of object is generated inside core region of the known CFL vortices, we simplified
the magnetic-induced symmetry breaking as $U(1)_{C+F}$ breaking. As the consequences, a superfluid-like
vortex string made up of the $\delta$-condensate was constructed and then the discussion was extended to the situation where not merely the $\delta$-condensate but also the $d$-condensate are incorporated. For the first time, a novel scenario
for vorton formation was suggested in the scope of magnetic color-flavor-locked matter of dense QCD.

In the present work, we only considered the ideal situation with a uniform, flavor-independent value for
color-superconducting gaps. In fact, the CFL gaps themselves need to be differentiated according to their
microscopic content. In view of that the diquark condensate made up of rotated-charge neutral quarks is
different from that of opposite-charged quarks, there exist distinctive magnetic responses for two species
of condensates. It leads to the splitting behavior of gaps, which is the another side of the same coin
regarding magnetic effect. The influences of gap splitting on the CFL vortices had been reported in our
previous work~\cite{zhang2015magnetic}. It is beyond the scope of current
paper since we focus on the exploration of the vortices consisting of MCFL.
%
More complexities come from the realistic situations. When a large strange-flavor mass is considered,
the color-flavor-locked symmetry becomes violated. In this case, CFL with $K$ condensate and the
resulting vortex solutions need to be included as well~\cite{kaplan2002charged,buckley2002superconducting}.

Another element missing in this work is rotated gluons originated from the rotated electromagnetic mechanism. Within the present GL framework the gluon dof with non-vanishing Meissner mass have been ignored for simplicity. 
In the situation of strong gauge coupling, it is reasonable to assume that the gluon fields by themselves did not affect the symmetry breaking induced by an external background field $B$. Once the rotated gluons are taken into account, nevertheless,  the corresponding color magnetic field could play the role on formation of non-Abelian CFL vortices. When the topological object consisting of CFL are regarded as color magnetic fluxes, for instance, the gluon interactions as well as the gluon-photon mixing had been found relevant for the stability and electromagnetic properties \cite{vinci2012spontaneous,eto2010instabilities,iida2005magnetic}.
In this case the current schemes of MCFL vortices, in particular the superfluid string, need to be further reexamined.
On the other hand, a rich physics lies in the interplay of rotated gluons with the applied field $B$.
By considering magnetic responses of gluons (through their rotated charges), this issue had been studied within gluon mean-field theory at high densities~\cite{ferrer2006magnetic}. 
Another kind of vortices with gluon condensate were predicted which are totally different from our concerned vortices with diquark condensates. Also, it was pointed out that the anti-screening effect is possible for very strong (external) magnetic fields. 
Together with the above-mentioned topics, further studies on influences of rotated gluons on the vortices consisting of magnetic color-flavor-locked matter are interesting undoubtedly. Towards this goal, it is an important task to deeply understand the implications of $\delta$ from microscopic viewpoint and bridge ``gaps" between the model-independent approach and the phenomenological quark/gluon model. We need to incorporate quark/gluon interactions with rotated photons in a more consistent manner. Also, the non-perturbative method rooted in full QCD might be necessary. Possible improvements in theoretical methods will be explored in future studies.

\section*{Acknowledgements}

This work was supported by National Natural Science Foundation of
China ( NSFC ) under Contract No. 10875058.


\end{document}